# Temperature and pressure evolution of the crystal structure of $A_x(Fe_{1-y}Se)_2$ (A = Cs, Rb, K) studied by synchrotron powder diffraction

V. Svitlyk[a], D. Chernyshov[a], E. Pomjakushina[b], A. Krzton-Maziopa[b], K. Conder[b]

V. Pomjakushin[c], V. Dmitriev[a]

[a]*Swiss-Norwegian Beamlines at European Synchrotron Radiation Facility, BP 220, 38043 Grenoble, France*

[b]*Laboratory for Developments and Methods, PSI, CH-5232 Villigen PSI, Switzerland*

[c]*Laboratory for Neutron Scattering, Paul Scherrer Institut, CH-5232 Villigen PSI, Switzerland*

**Abstract**

Temperature-dependent synchrotron powder diffraction on $Cs_{0.83}(Fe_{0.86}Se)_2$ revealed first order $I4/m$ to $I4/mmm$ structural transformation around 216°C associated with the disorder of the Fe vacancies. Irreversibility observed during the transition is likely associated with a mobility of intercalated Alkali atoms. Pressure-dependent synchrotron powder diffraction on $Cs_{0.83}(Fe_{1-y}Se)_2$, $Rb_{0.85}(Fe_{1-y}Se)_2$ and $K_{0.8}(Fe_{1-y}Se)_2$ ($y \sim 0.14$) indicated that the $I4/m$ superstructure reflections are present up to pressures of 120 kbar. This may indicate that the ordering of the Fe vacancies is present in both superconducting and non-superconductive states.

## 1. Introduction

Discovery of superconductivity in $Cs_x(Fe_{1-y}Se)_2$, $Rb_x(Fe_{1-y}Se)_2$ and $K_x(Fe_{1-y}Se)_2$ [1-3] is one more step on the way to comprehend the mechanism of superconductivity in the family of layered Fe-based compounds. Very recently it has been found that these compounds are magnetically ordered at room temperature and remain ordered in the superconducting state [4], this is rather unusual property of a superconductor. Another, also to a certain extent unique structural feature, is the ordering of the Fe vacancies in the FeSe superconducting layer, initially suggested in 1986 [5, 6] and recently confirmed by the single crystal synchrotron, neutron powder and laboratory X-ray single crystal diffraction studies [7, 8]. Ordering is manifested in a set of weak superstructure reflections that disappear on heating thus indicating an offset of the order-disorder transition [7]. An ordered pattern of the Fe vacancies in the layer that carry magnetism and also is responsible for superconductivity was considered as an important component of superconductivity [9, 10].

Different possible structural models of the Fe vacancies ordering were reported in literature. Structural model in the $I4/m$ symmetry with a $\sqrt{5} \times \sqrt{5} \times 1$ super cell and five times larger volume compared to the basic $I4/mmm$ unit cell, ThCr$_2$Si$_2$-type structure, dominates in the literature [8, 11-13]. Another reported vacancy ordering models correspond to a super cell with the q-vector q1 = (3/5, 1/5, 0) and also structures with coexistence of the vector q1 and the vector q2 = (3/4, 1/4, 0) [14]. The 5x5x1 super cell was also reported [15] with a further revision in a $\sqrt{5} \times \sqrt{5} \times 1$ super cell [8].

Orthorhombic *Pmna* structure with a $\sqrt{2}\times\sqrt{2}\times 1$ unit cell dimensions of the $ThCr_2Si_2$-type basic cell was reported to coexist with the *I*4/*mmm* and *I*4/*m* structures simultaneously [16]. However, a symmetry-based phenomenological analysis enumerating possible ordering patterns within unified scheme have not been reported; there has been also no agreement on the order of the transition, since structural features characteristic for both, 1st and 2nd order transitions, have been seen in the diffraction experiments [7, 13].

A number of pressure-dependent studies on the structures and ordering temperatures of the layered Fe-based systems were carried out [17-27]. Very recently, the first pressure-dependent studies on the conductivity of $Cs_x(Fe_{1-y}Se)_2$ revealed a suppression of superconductivity at pressures about 80 kbars [28]. However, studies of structural behavior as a function of pressure have not been yet done for the $Cs_x(Fe_{1-y}Se)_2$, $Rb_x(Fe_{1-y}Se)_2$ and $K_x(Fe_{1-y}Se)_2$ phases.

A numerous electronic structure calculations for these and related compounds appeared as well [29-35], not necessarily in agreement with the experimental data. In fact, it is not completely proper to put side by side a theoretical model obtained for a temperature of 0 K and a real room temperature structure as found in diffraction experiments, especially with a presence of structural disorder. The robust link and, at the same time, control parameter between the theory and practice can be a bulk modulus and thermal expansion coefficient, which can be obtained independently theoretically and experimentally. Corresponding experimental data have not been reported so far.

Therefore, in spite of a great interest to the new family of superconductors, there are still many questions opened related to the phase transition phenomena and basic physical properties. In order to close these gaps we have studied systematically pressure and temperature evolution of crystal structures of $Cs_x(Fe_{1-y}Se)_2$, $Rb_x(Fe_{1-y}Se)_2$ and $K_x(Fe_{1-y}Se)_2$ with help of synchrotron radiation.

Here the first results of the pressure-dependent X-ray powder diffraction on $Cs_{0.83}(Fe_{1-y}Se)_2$, $Rb_{0.85}(Fe_{1-y}Se)_2$ and $K_{0.8}(Fe_{1-y}Se)_2$ ($y \sim 0.14$) are presented. The data collected as a function of pressure have been parameterized in a form of equation of state thus giving first experimental estimates of the bulk moduli. We also show that pressure does not suppress the ordering of the iron vacancies even in the pressure range where no superconductivity is expected. We report a detailed temperature dependence of the order parameter for $Cs_{0.83}(Fe_{0.86}Se)_2$ responsible for the ordering of the vacancies in the iron sub-lattice; we show that the sample's decomposition could go in parallel with the phase transformation. Our experimental findings are complemented with the phenomenological consideration of the observed phase transformation to link together numerous and sometime contradictory results reported so far.

2. **Experimental procedure**

Single crystals of the $Cs_x(Fe_{1-y}Se)_2$, $Rb_x(Fe_{1-y}Se)_2$ and $K_x(Fe_{1-y}Se)_2$ samples studied in this work are the same as in the Ref. [7]. The crystals were grown using the Bridgeman method and the details of the experimental procedure are described in [2]. The actual composition for the Cs-based compound was obtained during the previous single crystal diffraction experiment and is equal to $Cs_{0.83}(Fe_{0.86}Se)_2$. Compositions for the Rb- and K-based samples were taken as nominal for alkali metals and Fe deficiency was assumed to correspond to the one found in the Cs-based sample, namely $Rb_{0.85}(Fe_{1-y}Se)_2$ and $K_{0.8}(Fe_{1-y}Se)_2$ ($y \sim 0.14$). For simplicity we use $Cs_{0.83}(Fe_{1-y}Se)_2$, $Rb_{0.85}(Fe_{1-y}Se)_2$ and $K_{0.8}(Fe_{1-y}Se)_2$ formulae throughout the text.

For temperature-dependent powder diffraction, single crystals of the $Cs_{0.83}(Fe_{1-y}Se)_2$ sample were finely ground and sealed in quartz capillaries in the glovebox under argon atmosphere. During diffraction experiment, temperature was changed in the 58 – 268°C temperature range with a 2°C step on heating (Fig. 1) and on cooling to check for possible structural hysteresis.

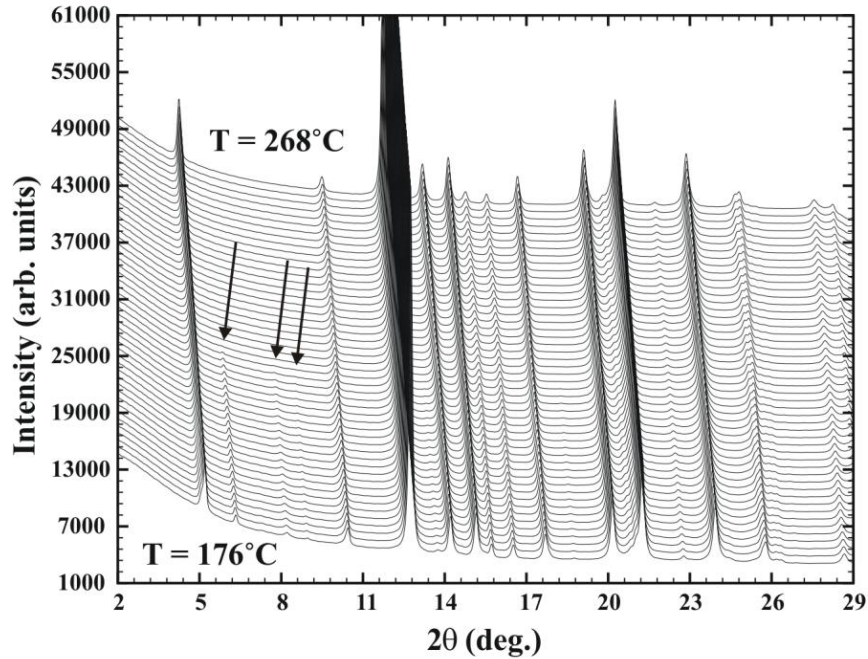

Figure 1. Powder diffraction data as a function temperature collected for the $Cs_{0.83}(Fe_{0.86}Se)_2$ sample (data in the 176 – 268°C temperature range is shown). The arrows indicate vanishing of the superstructure (110), (112) and (020) reflections upon heating, which are indicative of the Fe vacancies ordering

For pressure-dependent powder diffraction, single crystals of the $Cs_{0.83}(Fe_{1-y}Se)_2$, $Rb_{0.85}(Fe_{1-y}Se)_2$ and $K_{0.8}(Fe_{1-y}Se)_2$ samples were finely ground in the glovebox under argon atmosphere as well. Sealed samples were opened immediately before loading to the high pressure (HP) cells to minimize exposure to air. Silicon oil AP 100 was used as a pressure transmitting medium. A typical ethanol-methanol mixture, which retains good hydrostaticity at pressures up to 100 kbar, could not be used due to the high chemical reactivity of the samples. During diffraction experiments pressure was changed from 0 to 100-120 kbar, depending on the sample tolerance, with a typical step of 5 kbar on loading and 20 kbar on unloading.

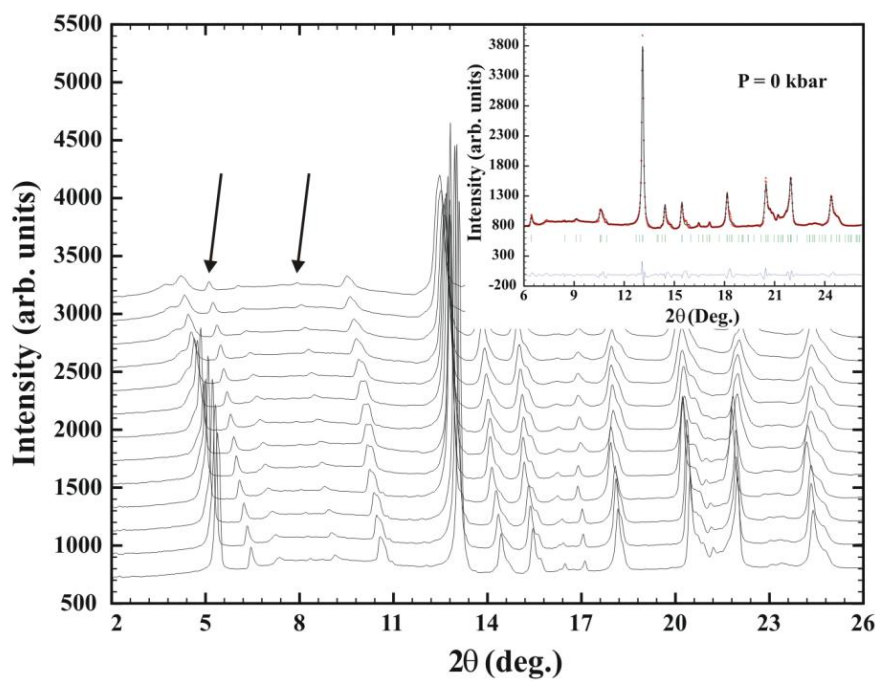

Figure 2. Powder diffraction data as a function of pressure, from ambient pressure (bottom) to 123 kbar (top) for the $Rb_{0.85}(Fe_{1-y}Se)_2$ sample. The arrows indicates (110) and (020) superstructure reflections. Insert shows refinement of the ambient pressure dataset within the $I4/m$ model

The best tolerance to the HP was exhibited by the $Rb_{0.85}(Fe_{1-y}Se)_2$ sample (Fig. 2), followed by the $Cs_{0.83}(Fe_{1-y}Se)_2$ and $K_{0.8}(Fe_{1-y}Se)_2$ samples. The poor hydrostaticity of silicon oil resulted in high induced strains in the samples at high pressures. As a result, powder data for all samples was treated in the average $I4/mmm$ structures. The Cs and Fe occupancies in the $Cs_{0.83}(Fe_{1-y}Se)_2$, $Rb_{0.85}(Fe_{1-y}Se)_2$ and $K_{0.8}(Fe_{1-y}Se)_2$ phases could not be reliably refined from the experimental powder data.

For all samples a peak of an unidentified secondary phase was seen at low angles close to the position of the 002 reflection (indexed in $I4/m$). The peak was disregarded and omitted from the refinement; most probably it comes from a secondary phase as a result of samples' surface degradation. Similar observation has been reported before [36] and prescribed to an inhomogeneous distribution of intercalated ions.

Both temperature- and pressure-dependent data collection was performed on the MAR345 diffractometer using a synchrotron radiation with the $\lambda = 0.69775$ Å. Data was integrated and processed with the FIT2D software [37, 38], standard deviations were calculated with a local script.

## 3. Results and discussion

### 3.1. Thermal expansion of $Cs_{0.83}(Fe_{1-y}Se)_2$

Dependence of the unit cell volume of $Cs_{0.83}(Fe_{1-y}Se)_2$ from temperature is presented on Figure 3, left (the data collected on heating is shown; in the paper graphs are presented without error bars when their size is the same or smaller than a size of the corresponding graphical markers). Anomaly around T = 216°C corresponds to the transition from the *I*4/*m* structure with the ordered Fe vacancies to the basic $ThCr_2Si_2$-type *I*4/*mmm* structure with the disordered Fe vacancies [7]. The obtained linear coefficients of thermal expansion, defined as $\alpha = \frac{1}{V_0} \cdot \frac{V-V_0}{T-T_0}$, where $V_0$ and $T_0$ are the initial unit cell volume and the sample temperature, respectively, $V$ and $T$ are the final unit cell volume and the sample temperature, respectively, for the low-temperature *I*4/*m* and high-temperature *I*4/*mmm* structures are $9.26 \cdot 10^{-5}$ and $9.93 \cdot 10^{-5}$ °C$^{-1}$, respectively. Some information on temperature evolution of the unit cell dimensions have been reported by us earlier [7], here we complement it with the unit cell ratio indicating rather anisotropic thermal expansion. One could see that all observed dependences agree with a 1$^{st}$ order character of the phase transition. Interestingly, there are broad hysteresis zones both above and below the transition temperature.

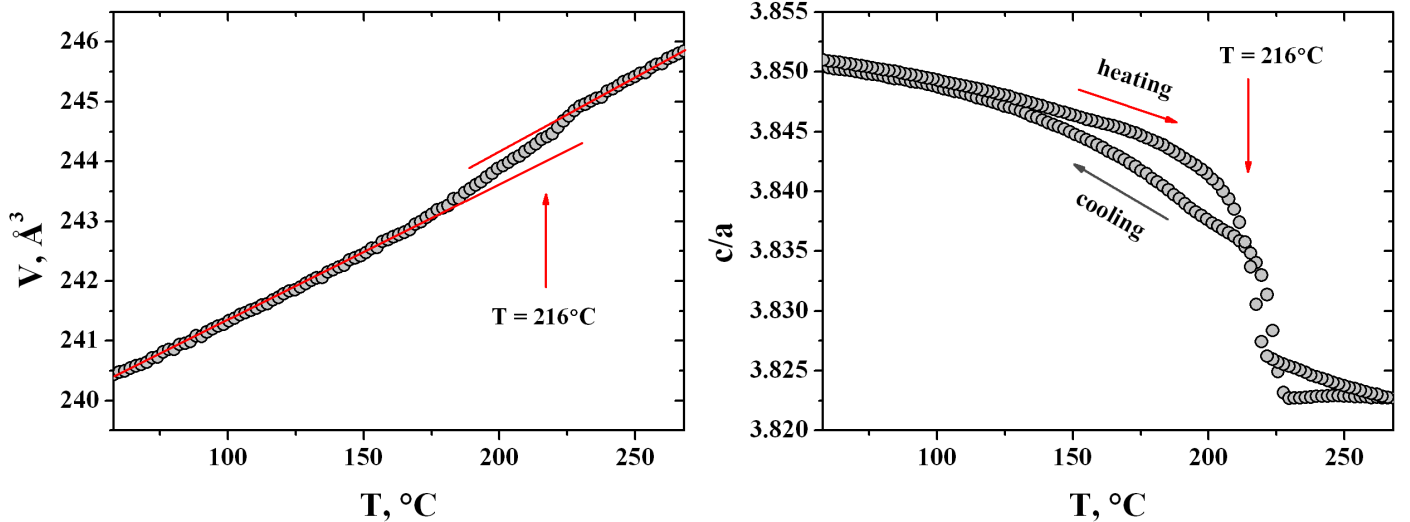

Figure 3. Dependence of the unit cell volume of $Cs_{0.83}(Fe_{1-y}Se)_2$ (left) and *c*/*a* parameters ratio (right) on temperature

## 3.2. Equations of state and bulk moduli for $Cs_{0.83}(Fe_{1-y}Se)_2$, $Rb_{0.85}(Fe_{1-y}Se)_2$ and $K_{0.8}(Fe_{1-y}Se)_2$

Volume vs. pressure dependences for the $Cs_{0.83}(Fe_{1-y}Se)_2$, $Rb_{0.85}(Fe_{1-y}Se)_2$ and $K_{0.8}(Fe_{1-y}Se)_2$ samples (*I*4/*mmm* setting) fitted with the Murnaghan equation of state (Eq. 1), where $V_0$ is the volume at zero pressure, $B_0$ is the bulk modulus and $B_0'$ is the first pressure derivative of the bulk modulus are shown on Fig. 4. Fitted parameters are summarized in Table. 1.

$$P(V) = V_0(1 + B'_0 \frac{P}{B_0})^{-\frac{1}{B'_0}} \quad (1)$$

Table 1. Experimental coefficients of the Murnaghan equation of state for $Cs_{0.83}(Fe_{1-y}Se)_2$, $Rb_{0.85}(Fe_{1-y}Se)_2$ and $K_{0.8}(Fe_{1-y}Se)_2$ from pressure-dependent synchrotron X-ray powder diffraction (*I4/mmm* settings)

| sample | $V_0$, Å$^3$ | $B_0$, kbar | $B_0'$ |
|---|---|---|---|
| $Cs_{0.83}(Fe_{1-y}Se)_2$ | 241.6(3) | 223(10) | 5.8(3) |
| $Rb_{0.85}(Fe_{1-y}Se)_2$ | 227.7(7) | 260(24) | 5.9(7) |
| $K_{0.8}(Fe_{1-y}Se)_2$ | 209(2) | 335(33) | 5.8 (fixed) |

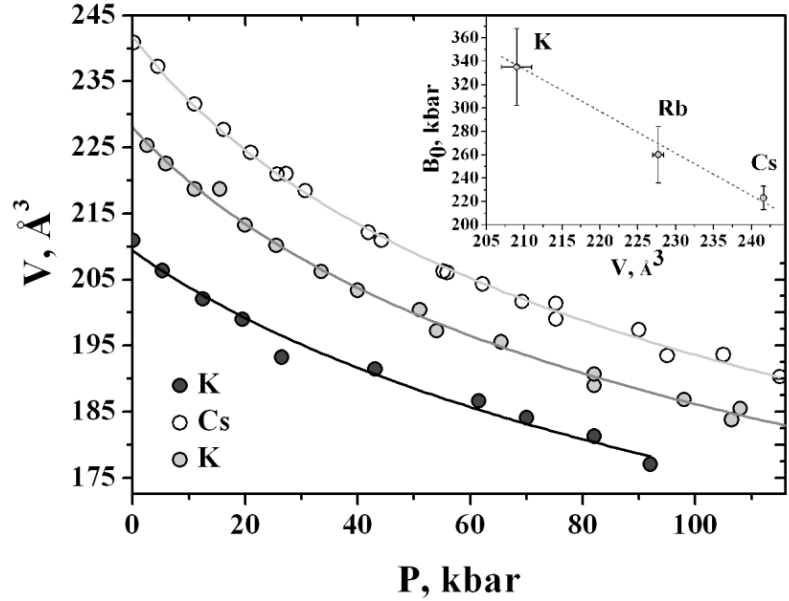

Figure 4. Volume vs. pressure dependences for $Cs_{0.83}(Fe_{1-y}Se)_2$, $Rb_{0.85}(Fe_{1-y}Se)_2$ and $K_{0.8}(Fe_{1-y}Se)_2$ fitted with the Murnaghan equation of state. Insert shows bulk moduli versus the unit cell volumes at ambient pressure

Since ionic radius of alcali ions increase from K to Cs, observed increase of the $V_0$ is expected. A decrease of the bulk modulus with ionic radius increasing is less obvious and may be even considered contrintuitive. It is, however, also expected if we account for the stoichiometry - by adding bigger alkali ions we also create larger vacancies' voids in this sublattice, thus making structure less dense and therefore more compressive. As a result, we observe a negative correlation between the bulk moduli and unit cell volumes at ambient pressure (Fig. 4, inset).

### 3.3. Temperature and pressure-dependent crystallographic behavior of $Cs_{0.83}(Fe_{1-y}Se)_2$, $Rb_{0.85}(Fe_{1-y}Se)_2$ and $K_{0.8}(Fe_{1-y}Se)_2$

First, we should note a lack of complete reversibility in temperature and pressure scans. Temperature-dependent anomalies related to the order-disorder phase transition could be observed starting at 216°C on heating (Fig. 5). The most prominent changes were exhibited by $c$ cell parameter, Se $z$ coordinate and, correspondingly, by the height of Se atoms (Fig. 5, right). The latter parameter is considered as important in structure-property correlations for the Fe-based superconductors [39]. While hysteresis indicates a first order of the transition near 216°C, irreversibility of the temperature evolution might be related to a certain mobility of Cs cations, even within sealed capillary. One, therefore, has to avoid diffraction experiments under dynamic vacuum to prevent hardly predictable changes in a stoichiometry.

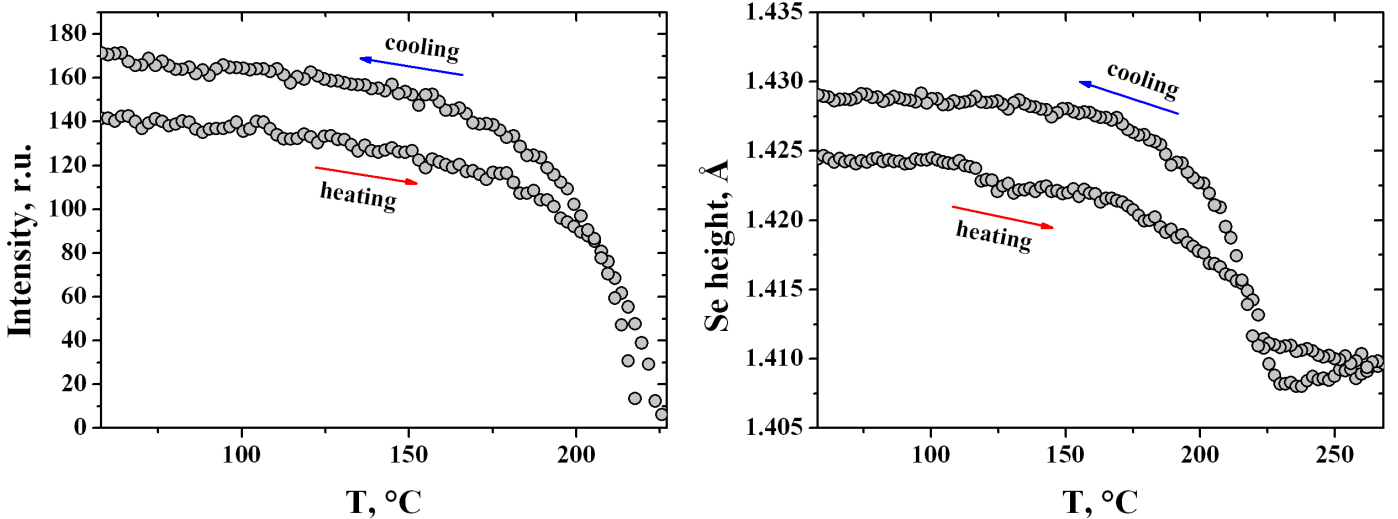

Fig. 5 Temperature dependence of intensity of the 110 superstructure reflection (*I*4/*m*) (left) and the height of Se atoms in the FeSe layer (right, errors average is equal to 4·10$^{-3}$) in $Cs_{0.83}(Fe_{1-y}Se)_2$

Dependence of the anion height from pressure can be used as a convenient tool to track possible changes in the conductivity of samples [39]. In our case no pronounced pressure-dependent anomalies have been found for the $Cs_{0.83}(Fe_{1-y}Se)_2$, $Rb_{0.85}(Fe_{1-y}Se)_2$ and $K_{0.8}(Fe_{1-y}Se)_2$ samples. Big spread of experimental datapoints was observed which originates from the correlations between structural parameters unavoidably appearing due to the quality of diffraction patterns limited by high-pressure experiment.

The $I4/m$ superstructure peaks did not disappear on applying the pressure in the $Cs_{0.83}(Fe_{1-y}Se)_2$, $Rb_{0.85}(Fe_{1-y}Se)_2$ and $K_{0.8}(Fe_{1-y}Se)_2$ samples. This indicates that the ordering of the Fe vacancies was not suppressed within the range of applied pressures.

### 3.4. Phenomenological analysis of order-disorder phase transition

Analysis of the possible structural transformations in $A_x(Fe_{1-y}Se)_2$ following a symmetry based phenomenological scheme is needed to validate the experimentally obtained models. Accurate single-crystal diffraction experiments [7] show that the structural distortion, which is the ordering of the iron vacancies, propagates with the reciprocal space vector $\mathbf{k}_{2i}=[2/5, 1/5, 1]$ which is identical to $\mathbf{k}_{2p}=[2/5, -2/5, 4/5]$ and corresponds to the Θ-point of the tetragonal Brillouin zone (BZ). We use hereafter reciprocal k-vectors both in the notations of the International Tables for Crystallography [40] (denoted by index "i") dominating in the experimental works, and in the primitive basis according to [41], which is mainly conventional in the group-theoretical exercises. Basis transformation from the parent Bravais lattice (lowercase letters) to the distorted one reads:

$\mathbf{A}=2\mathbf{a}+\mathbf{b}$; $\mathbf{B}=-\mathbf{a}+2\mathbf{b}$; $\mathbf{C}=\mathbf{c}$.         (2)

Vectors in (2) identify two body-centered tetragonal Bravais cells. Orientational relationships between the corresponding primitive (independent) unit cells look as following:

$\mathbf{d}_1=\mathbf{a}_1-\mathbf{a}_2-\mathbf{a}_3$, $\mathbf{d}_2=2\mathbf{a}_2+\mathbf{a}_3$, $\mathbf{d}_3=\mathbf{a}_1+2\mathbf{a}_3$.         (3)

The basal vectors (**A**, **B**, **C**) of the distorted tetragonal structure (2) are not parallel neither to the same vectors (**a**, **b**, **c**) in the parent lattice nor to its diagonal directions. It should result in the loss of the corresponding mirror planes parallel to and rotation axes perpendicular to the four-fold axis in the distorted structure. A maximal crystal class for the low-symmetry structure can be therefore indentified as $4/m$. Taking into account the systematic absences conditions, indicating a body-centered lattice for the low-symmetry phase, the space group $I4/m$ is suggested. Among other candidates $P4_2/n$ with the same Bravais cell parameters (2) could be considered. However, the corresponding phase transformation $I4/mmm$-$P4_2/n$ (unit cell volume multiplication V'/V=10) involves, in addition to the above mentioned distortion propagating with the $\mathbf{k}_2$, one more order parameter characterized by the $\mathbf{k}_{15i}=[0, 0, ½]$. Careful investigations conclude the absence of the corresponding reflections violating $I$-centering in the diffraction patterns of the studied compounds.

Another candidate, already appeared in the earlier publications but latter rejected [8, 15], is the $I4/mmm$ structure with the unit cell volume multiplied by a factor of 25 with respect to the parent structure. Basis transformations between the crystal lattices are

**A'**=5**a**; **B'**=5**a**; **C'**=**c**         (4)

indicating that all eight arms of the vector star $\mathbf{k}_2$ participate in the propagation of the distortions during the $I4/mmm$-$I4/mmm$ (V'/V=25) phase transition. It should manifest, in turn, in identical intensity of all eight new equal satellites of the ±[2/5 1/5 1] type. In contrast, the superstructure in the $I4/m$ phase is formed by four out of eight star vectors: $\mathbf{k}_{2p}^{(1,2)}= \pm[2/5, -2/5, 4/5]$ and $\mathbf{k}_{2p}^{(3,4)}= \pm[4/5, 1/5, -2/5]$, and the condition of the equal intensities is satisfied in two quartets of the reflections, each corresponding to the different system of translational (antiphase, in other terminology) domains [7, 8].

For the sake of completeness, it is worth mentioning an orthorhombic structure found by the authors in Ref. [16]. An additional phase described by the space group *Pmna* with the supercell of $\sqrt{2}$x$\sqrt{2}$x1 of the tetragonal parent $I4/mmm$ was reported to exist in some samples between 295 and 500 K. Our analysis concludes that the corresponding structure distortion should propagate with the $\mathbf{k}_{13i}=[½, ½, 0]$. An independent structure order parameter belongs to the X-point [½ ½ 0] of the tetragonal BZ. Moreover, temperature treatment of the corresponding samples proves a metastable character of the *Pmna* structure, which is sensitive to the sample preparation conditions. The orthorhombic phase never appeared in our syntheses, and only single-phase sample were used in the present experiments.

## 4. Conclusions

Temperature ($Cs_{0.83}(Fe_{1-y}Se)_2$) and pressure evolution of crystal structures ($Cs_{0.83}(Fe_{1-y}Se)_2$, $Rb_{0.85}(Fe_{1-y}Se)_2$ and $K_{0.8}(Fe_{1-y}Se)_2$) were studied using synchrotron powder diffraction. The structure of $Cs_{0.83}(Fe_{1-y}Se)_2$ possesses the phase transition on heating related to the disorder of the iron vacancies in the FeSe layer, as we have shown before and in agreement with other reports [7, 16]. At variance with the results reported in [14], we found the transition to be of the first order as evidenced by hysteresis in the lattice properties and also agrees with the differential scanning calorimetry [7]. At the same time, we have noted an irreversibility of the temperature dependence of the lattice dimensions and Bragg intensities. We tentatively relate the irreversibility to a mobility of the intercalating alkali ions; such a mobility may seriously affect apparent temperature evolution of the crystal structure.

Our symmetry-based phenomenological analysis enumerates crystal structures reported so far. Among them, $I4/mmm$ can be considered as the parent one, and two possible structures are associated with the vector star $\mathbf{k}_2$, $I4/m$ and $I4/mmm$. Formation of the hypothetic $P4_2/n$ structure involves, in addition to the above mentioned distortion propagating with the $\mathbf{k}_2$, one more order parameter characterized by the $\mathbf{k}_{15}$ wave vector. Orthorhombic *Pmna* structure with a $\sqrt{2}$x$\sqrt{2}$x1 unit cell dimensions can be obtained from the parent one by the distortion associated with the $\mathbf{k}_{13}$ wave vector; a coexistence of this phase with the parent $I4/mmm$ and $I4/m$ ($\mathbf{k}_2$) reported in [16] is rather unusual and has to be confirmed by single crystal diffraction experiments. Symmetry imposes no constrains on the order of the transition between parent $I4/mmm$ and $I4/m$ ($\mathbf{k}_2$), $I4/mmm$ ($\mathbf{k}_2$), $P4_2/n$ ($\mathbf{k}_2$, $\mathbf{k}_{15}$) structures; our experimental data indicates rather pronounced 1$^{st}$ order transition between $I4/mmm$ and $I4/m$ ($\mathbf{k}_2$) structures. One has, however, to take the temperature evolution of the structure and, especially, order parameter carefully, since the critical dependence may be strongly affected by the temperature variation of the sample's composition. Our experiment shows that such variation takes place even in a sealed capillary; the effect is expected to be much more pronounced if a sample has been heated under vacuum during data collection.

In spite of numerous experimental problems related to sample composition, homogeneity, and reactivity, our diffraction experiments as a function of pressure allowed to recover equation of states and, for the first time, provide with the experimental estimates of the bulk moduli. Together with the thermal expansion coefficients calculated from the temperature dependent diffraction, these data may serve as constrains for theoretical models. We have also seen that the Bragg reflections indicative of the ordering of the Fe vacancies in the FeSe layer do not disappear up to ~120 kbar. No clear anomaly for $Cs_{0.83}(Fe_{1-y}Se)_2$, $Rb_{0.85}(Fe_{1-y}Se)_2$ and $K_{0.8}(Fe_{1-y}Se)_2$ were observed around 80 kbar where transport experiments indicate an offset of superconductivity; this conclusion, however, may be limited by quality of the data. Albeit definitive answer can only be obtained from diffraction experiment at high pressure and very low temperatures, our data assume that the ordering of the vacancies in the FeSe layer may be present in both superconducting and non-superconducting states.

**Acknowledgments:** A. Krzton-Maziopa acknowledges the support by the Scientific Exchange Programme Sciex-NMSch (Project Code 10.048) and E. Pomjakushina acknowledges the support by the NCCR MaNEP Project.